\documentclass{llncs}
\usepackage{color}
\usepackage[dvipsnames]{xcolor}
\usepackage{mathrsfs,amsmath}
\usepackage{amsfonts}
\usepackage{graphicx}
\usepackage[title]{appendix}
\usepackage[noend]{algpseudocode}
\usepackage{algorithm}
\usepackage{algpseudocode}

\makeatletter
\renewcommand{\ALG@name}{Pseudocode}
\makeatother

\newfloat{protocol}{htbp}{loa}
\floatname{protocol}{Protocol}
\makeatletter\newcommand\l@subroutine{\@dottedtocline{1}{1.5em}{2.3em}}\makeatother

\usepackage{subfigure}
\usepackage[caption=false]{subfig}

\usepackage{enumitem}
\usepackage{color}
\usepackage[dvipsnames]{xcolor}

\usepackage[switch]{lineno}

\usepackage{cancel}

\begin{document}

\title{How Does Knowledge Come By?}
\titlerunning{Hamiltonian Mechanics}  
%
\author{Anamika Chhabra \and S.R.S. Iyengar}

\institute{Indian Institute of Technology Ropar, India\\    \email{anamika.chhabra@iitrpr.ac.in, sudarshan@iitrpr.ac.in}}
%
%

\maketitle              

\begin{abstract}
Although the amount of knowledge that the humans possess has been gradually increasing, we still do not know the procedure and conditions that lead to the creation of new knowledge. An understanding of the modus operandi for the creation of knowledge may help in accelerating the existing pace of building knowledge. Our state of ignorance regarding various aspects of the process of knowledge building is highlighted by the existing literature in the domain. The reason behind it has been our inability to acquire the underlying data of this complex process. However, current time shows great promise of improvements in the knowledge building domain due to the availability of several online knowledge building portals. In this report, we emphasize that these portals act as prototypes for universal knowledge building process. The analysis of big data availed from these portals may equip the knowledge building researchers with the much needed meta-knowledge. 
\end{abstract}


%
%

\section{Introduction}
It is true that we have relatively progressed with time in terms of the amount of knowledge that we have gathered about the things around us. For example, our current knowledge of physics and quantum mechanics is satisfactorily able to answer many of our queries. Nonetheless, it is terrible to make a note of all those things that we are still clueless about. For instance, we are still unsure about how this universe came into existence. Further, is there any other parallel universe out there or rather we are a tiny element of a gigantic multiverse? Even after so many years of inquisitiveness, we still do not know whether life exists on other planets. Till date, we are not able to find cure for a large number of fatal diseases. Many of the mathematical conjectures that we ourselves developed, we have not been able to prove them. All these instances, which are a representative subset of a larger set of unknown things, indicate that our current knowledge base is not sufficient to answer many important questions. Additionally, an existing study indicates that an enormous part of world's knowledge is still unexplored \cite{pennisi2003modernizing}, which brings our attention to focus on the need to expand our existing knowledge base.


The preceding discussion indicates that the pace with which we are building new knowledge and the methods we have been adopting so far are not satisfactory and hence require attention. However, in order to bring improvements in the way we are building knowledge, we first need to answer the question as to \textit{how new knowledge comes by}. An analogy would be, to build better spacecrafts, we first need to know how we built the existing spacecrafts or in general, how spacecrafts are built. Unfortunately, we still do not know the answer to this very fundamental question i.e. how knowledge gets built. The knowledge that we have gathered till date seems to have been built out of some serendipitous methods or from techniques based on trial and error. Understanding which situations or parameters lead to the creation of new knowledge requires one to go one level deeper and try to perceive the underlying activities of the process. In other words, for a new piece of knowledge built, it requires one to thoroughly analyze the sequence of actions that took place in building this piece, in a reverse fashion. This kind of analysis, therefore, necessitates the acquisition of the footprints of knowledge building. 


Apart from our inability to answer the causes for creation of new knowledge, a study of the literature points to the existence of a number of critical questions pertaining to the development of knowledge. These questions are either unanswered or in a state of debate. Finding an answer to these questions is imperative to understand the broad question of how new knowledge gets built. For example, one of the controversies is whether people making small contributions are essential in a knowledge building environment or they are easily replaceable without any damage to the system? This leads to the lack of consensus over the validity of two contradictory hypotheses named \textit{Ortega Hypothesis} \cite{cole1972ortega}\cite{macroberts1987testing} and \textit{Newton Hypothesis} \cite{bornmann2010scientific}. This important subject matter still remains in a state of disagreement where Ortega Hypothesis gives due regard to a mass of people making small contributions whereas Newton Hypothesis believes that a large mass of people making small contributions can be harmlessly removed from the system. Another uncertainty is in the context of the number of participants for efficient knowledge building. Given that less number of people bring less perspectives into the system and a lot of people lead to increase in conflicts and co-ordination costs, then what is the ideal number of participants for building knowledge? Further, we have seen that diversity is important for knowledge building \cite{hong2004groups}, but at the same time too much diversity gives rise to problems such as lack of consensus \cite{chen2010effects}? Therefore, what is the right amount of diversity for knowledge building? These are just a few examples of some of the huge gaps in our understanding of the domain. Section 3 highlights these and many other research questions that are still unanswered in the domain. To be able to answer the broad question of how new knowledge gets built, it is imperative to have a clear understanding of these associated aspects of the process of knowledge building. Further, interpreting the different aspects of the domain will entail obtaining the underlying data that could keep track of the background activities of the process.

Building new knowledge is a complex phenomenon involving a number of social, psychological and cognitive processes. It also encompasses a number of abstract and intangible activities that are indeed difficult to record. Therefore, getting the appropriate data for understanding how new knowledge gets built and the associated aspects of this general question is a challenging task. Unfortunately, our inability to have an access to such data has resulted in our limited understanding of the domain as highlighted by the examples just discussed. However, since getting data from the actual environment is difficult, we may try getting data from the virtual knowledge building environments. We, therefore, discuss alternate ways and techniques that may provide us the underlying data to understand various aspects of the phenomenon. 

\begin{figure}
\centering
\includegraphics[scale = 0.35]{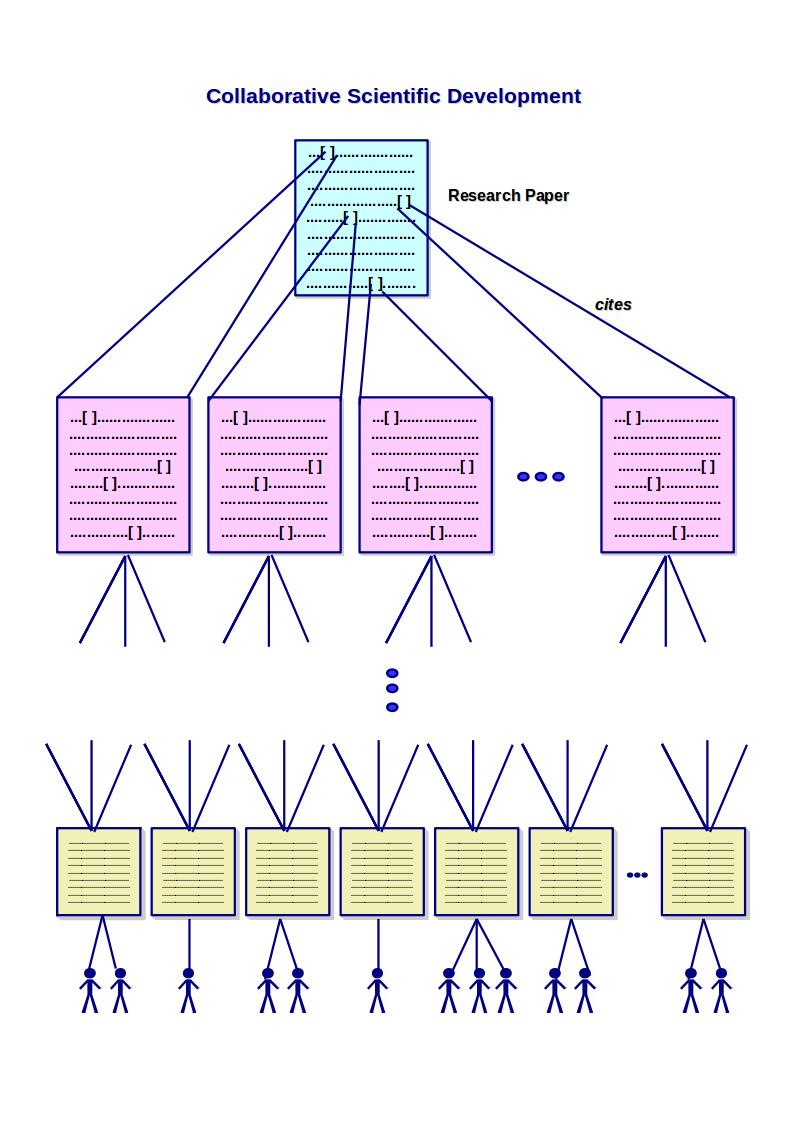}
\caption{Collaborative Scientific Developments}
\label{collaborative_citation}
\end{figure}
 

\subsection{Can Citation Data sets help?}
Let us first attempt exploring a domain that maintains a record of a subset of the universal knowledge, i.e the domain of scientific developments. The published papers in journals and conferences provide a documentation of the advancements in science. Therefore, the data set from this domain might facilitate a way to decipher the scientific knowledge building process. Let us briefly take a look at how scientific knowledge development takes place. A research paper, although having a fixed number of authors, is broadly the result of a collaborative effort of a large number of people, including the authors of the paper, the authors of the papers that a paper cites, along with many other un-cited people. Further, this collaboration takes place in various steps leading to a recursive knowledge development (See Figure~\ref{collaborative_citation}. As part of the collaboration, a paper may take help from another paper by building on top of it or it may prove the approach used by another paper to be wrong or it may correct the approach of another paper by providing another approach. Therefore, a particular paper providing evidence of a piece of scientific knowledge advancement is not an outcome of the efforts of just the set of authors listed on the paper, rather it is a gradual process of building knowledge on top of the knowledge built by many others. It will be difficult to come across an invention that does not make use of the existing knowledge. Since the universal knowledge building process is also known to be collaborative, studying the citation data of the published papers may provide us insights into the dynamics of how new scientific knowledge gets built collaboratively. This citation data may be taken from various bibliographic databases such as Scopus, Web of Science and Google Scholar. However, there are two limitations over the use of this data to achieve our goals. Firstly, apart from providing some quantitative details about who cites whom over a time line, the citation data sets do not provide any other in-depth details of the phenomenon. Secondly, the literature shows serious objections over the studies that have been conducted just on the basis of citation data. This is due to the problems that have been observed in the citation practices of the researchers. Some of these problems include the authors making ceremonial citations to friends, colleagues or eminent people in the field \cite{cole1972ortega}\cite{meho2007rise}, taking citations from another paper and adding them to their own paper without reading them, i.e. \textit{Secondary Citing} or \textit{Tertiary Citing} \cite{hoerman1995secondary}, an author using different names, i.e. \textit{Author Disambiguation} or different authors having same names, the rich-getting-richer phenomenon, i.e. \textit{Matthew Effect in Citations} \cite{wang2014unpacking}\cite{merton1968matthew}, \textit{Halo Effect} \cite{cole1972ortega} and not citing old papers assuming them to be well-known standards, i.e. \textit{Immediacy Factor} etc. This discussion indicates that the citation data set may provide us only limited details about the process and prompts us to look for some alternative ways to comprehend deeper aspects of the process of knowledge building.


\begin{figure}
\centering
\includegraphics[scale = 0.30]{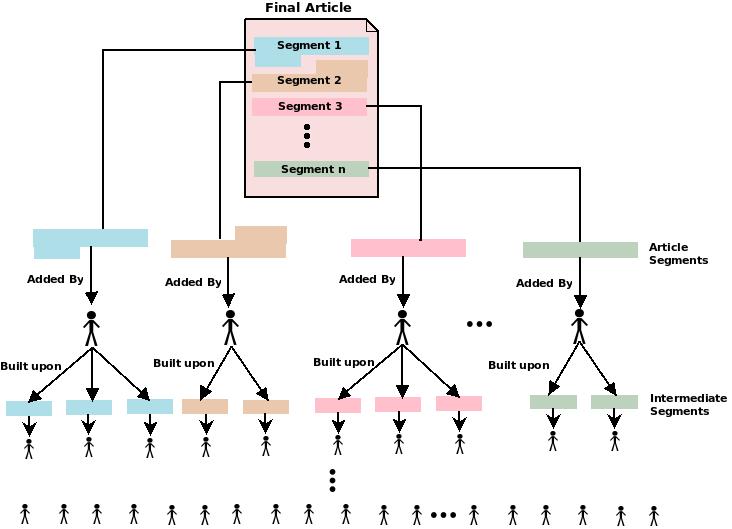}
\caption{Collaborative Development of Knowledge in a Wikipedia Article}
\label{collaborative_wikipedia}
\end{figure}

\subsection{Can the data sets from online knowledge building portals help?}
The advancements in the Internet technologies have facilitated collaboration at an unprecedented scale by providing several \textit{crowdsourced} portals. These portals are successfully accomplishing the job of collective problem solving, collaboratively accumulating facts and building knowledge on top of the existing information. Some of the examples are the online openly editable encyclopedia Wikipedia and Q\&A portals such as Stack Exchange and Quora etc. We conjecture that these portals emulate various aspects of the universal knowledge building process. As an example, Figure~\ref{collaborative_citation} depicts how knowledge in Wikipedia articles is built collaboratively. Every segment of the final article is a result of a sequence of operations over the content previously added by a bunch of other contributors. This process keeps going on recursively where each contributor takes the article from one state to the other. Overall, the content of the article at any point of time is the result of collaboration between a large number of contributors. Wikipedia contributors may insert new content or they may delete or modify the content added by others. This is analogous to a paper providing an \textit{adjacent next} to the technique of another paper or refuting or improving upon the hypothesis provided by another paper respectively in a research environment. The similarity between Figure~\ref{collaborative_citation} and \ref{collaborative_wikipedia} indicates that an analysis of the data set of Wikipedia might help us uncover many aspects of collaborative knowledge creation. That is, if we carefully observe how an article that starts with a rough first paragraph takes the shape of the best-known resource on any given topic gradually, we might be able to answer some of the long-unresolved queries regarding collaboration. This analogy indicates Wikipedia emulating one aspect, i.e. collaboration of the knowledge building process. In a similar way, many other portals such as Q\&A portals can help us understand other aspects of the phenomenon. Although these portals differ in their features, they all aim at building a repository of the world's knowledge by a collective effort of the crowd. Further, the diversity in the approaches employed by these crowdsourced mediums is advantageous in understanding different characteristics of the process. For instance, while Wikipedia may help in uncovering the aspects of co-ordination and conflicts while building repositories of knowledge, Q\&A portals may assist in deciphering the importance of discussions and incentivization etc. Additionally, these portals are able to store each and every footprint of the process digitally, which is otherwise not possible to acquire. For instance, every single edit that has transformed a Wikipedia article - which started out as a few badly written sentences on a blank `slate' and has now emerged out as a great artifact of knowledge about a particular topic - has been recorded by the portal and kept open in public for analysis. The meta-data of every question and answer along with its votes, comments, edits and the timing details is documented in an easy-to-analyze form by Stack Exchange and is made available regularly. These features make the crowdsourced portals ideal for understanding various aspects of knowledge building. 

With so many platforms enabling an unprecedented collaboration among people from every nook and cranny of the universe, we may get a huge amount of data for exploration. We, therefore, envision that proper analyses conducted on the big data taken from these portals may shed light on deciphering how new knowledge gets created. These analyses may further help in uncovering several other characteristics of collaborative knowledge creation facilitating unprecedented improvements in the domain. Therefore, just the way mice are used to test human medicines to examine their side-effects, we suggest use of these portals for understanding how knowledge building happens in general. The forthcoming sections provide the details of the current scenario of the knowledge building domain along with the challenges associated with the realization of the proposed vision.

\section{Current Scenario}
The proposed vision makes an attempt to bring improvements in the domain of \textit{knowledge building} by using the data sets from crowdsourced portals, which come under the \textit{computer science} domain. This section outlines the current scenario of both the domains. It first describes the work done in the knowledge building domain. Then it reports the work attempted in the computer science domain, i.e. the analyses performed on the data sets of the portals. Unfortunately, the literature indicates a limited amount of work attempted in the knowledge building domain, that too is either theoretical or performed on very small scale analyses. On the other hand, computer science researchers have been analyzing the data sets from crowdsourced portals to a decent extent. However, these analyses have been primarily focused on either understanding or improving some portal specific feature, rather than on understanding the universal knowledge building process. In the end, this section reports the small amount of analyses made on the data sets to uncover some knowledge building aspect. 

\subsection{Knowledge Building Work}
Nonaka \cite{nonaka1994dynamic} in 1994 stated that new knowledge gets created through a constant dialogue between \textit{tacit knowledge}\footnote{It is codified knowledge which is transmittable in some formal language} and \textit{explicit knowledge}\footnote{It is the knowledge that human beings possess and acquire through their experience over time}. He developed a theoretical framework named \textit{SECI model} that provides an analytical perspective on various dimensions of knowledge creation. The model reports that knowledge creation is about continuous transfer, combination, and conversion of the different types of knowledge, as users practice, interact, and learn. Although knowledge is being continuously built from the time immemorial, the formal definition to the term was given by Scardamalia and Bereiter \cite{scardamalia1994computer} in 1994. They further argued that the process of learning and the process by which the knowledge advances are quite similar. They emphasized the importance of a collaborative effort rather than individual effort for accelerated knowledge building. Cook and Brown \cite{cook1999bridging} in 1999 suggested that knowledge creation is a product of the interplay between knowledge and \textit{knowing}. The authors asserted that there is a difference between the possession of knowledge and the act of knowing. Knowing is something that comes about through practice, action and interaction. Further, the shift from knowledge to knowing is basically the driving force behind the creation of new knowledge. Gerry Stahl \cite {stahl2000model} provided a conceptual framework for collaborative KBEs in 2000 that consists of important phases that should be supported by any computer-based KBE. The model presented by the author explains the relationship of collaborative group processes to individual cognitive processes. All these works attempted towards understanding the causes of knowledge creation are mostly theoretical and limited in terms of their validation and general applicability. This is primarily due to the lack of underlying data or any medium through which the phenomenon could be understood. 

\subsection{Work Done using data sets of Knowledge Building Portals}
This subsection briefly outlines some of the work done on the data sets obtained from crowdsourced knowledge building portals.

In the context of Wikipedia, Viegas et al. \cite{viegas2004studying} developed a tool called \textit{History Flow Visualization} to understand the collaboration patterns of contributors of articles on Wikipedia. The tool uses the edit history of a given article to display a visualization of the events that took place while shaping the article. The paper also shows that the cases of vandalism are quickly handled by the Wikipedia community. A study published in nature indicates that the quality of Wikipedia articles is equivalent to the encyclopedia \textit{Britannica}, which is written by experts \cite{giles2005internet}. Suh et al. \cite{suh2007us} described a tool called `\textit{RevertGraph}' to visualize the reverts that take place on Wikipedia articles. Arazy et al. \cite{arazy2010recognizing} identified five activities performed by participants of Wikipedia viz. add (sentence), delete (sentence), proofread (word modification), improve navigation (internal hyperlinks) and link (hyperlinks to other wiki pages), whereas, Pfeil et al. \cite{pfeil2006cultural} identified thirteen categories that describe the actions performed by the contributors viz. add information, add link, clarify information, delete information, delete link, fix link, format, grammar, mark-up language, reversion, spelling, typography and vandalism. Similarly, Ram et al. \cite{liu2011does} categorized Wikipedians' actions into ten categories and tried to find their collaboration patterns. Kittur et al. \cite{kittur2008can} reported that transparency and visualization that exposes various aspects of the articles can significantly improve the trust of users in a mutable system such as Wiki. After a study on Wikipedia articles, Cress et al. \cite{cress2008systemic} observed various processes of learning and collaborative knowledge building through Wikis. However, the analysis that they conducted has been performed only on a few articles of Wikipedia, which is too small to be generalized and requires large scale verification. 

Anderson et. al \cite{anderson2012discovering} made use of the temporal structure of StackOverflow to identify which questions and answers are going to be of lasting value, and which of these require more attention from the contributors. The authors also observed that the high reputation users are mostly the ones who answer a lot and do not ask many questions. Further, most high reputation users acquired high reputation because of their answers getting accepted rather than due to upvotes. Movshovitz-Attias et. al \cite{movshovitz2013analysis} analyzed the Stack Overflow reputation system and observed that the users can be divided into high and low reputation users according to their contribution. It was confirmed that while most of the answers come from the high reputation users, most of the questions are asked by low reputation users. However, on an average, a high reputation user asks more questions than a low reputation user. The authors also predicted whether a given user will later turn into a significant contributor or not. Adamic et. al \cite{adamic2008knowledge} analyzed the Yahoo Answers knowledge sharing dynamics. One of the findings of their study was that lower entropy of their contribution across categories correlates with a higher probability of them receiving higher answer ratings. Wang et al. \cite{wang2013wisdom} studied three constituent networks of Quora, i.e. the network connecting topics to users, the network connecting users, and the network connecting related questions. The authors showed that heterogeneity in the user and question networks is one of the reasons behind the success of the portal. 

Cranshaw and Kittur \cite{cranshaw2011polymath} examined the \textit{Polymath Project}, which is a portal for collaboration amongst mathematicians and observed the role of leadership in the success of the portal and the importance of newcomers. Vasilescu et al. \cite{vasilescu2013stackoverflow} used the data from StackOverflow and Github and observed the correlation between the activities of users on both the portals. The authors observed that people who are very active on Github provide more answers on StackOverflow and ask very few questions. Similarly, people who ask a lot of questions show non-uniformity in their work on Github. It can be seen that most of the work on the data sets focuses on the portal specific details rather than on understanding the phenomenon of building knowledge, i.e. the work is on knowledge building portals and not on knowledge building process. We, therefore, find the usage of data sets quite under-explored in the context of knowledge building and perceive it as a prospective domain for the knowledge building researchers to investigate.

\section{Knowledge Building Dynamics: A Conundrum}
If we somehow are able to quantify the amount of knowledge that the world possesses today, it will come out to be way more than the knowledge we had a few decades ago. This is primarily due of the development of various tools enabling easy collaboration resulting into the swift creation of new knowledge. An important point to note is that although many of us are participating in the development of knowledge, we hardly know how we are doing it and what leads to the creation of knowledge. Understanding this process is important for the knowledge building domain to progress and to accelerate the pace of building knowledge. Further, not only do we not know how new knowledge comes by, we are in a state of ignorance when it comes to answering some of the related questions of this broad phenomenon. Answers to these questions will be required for us to be able to understand the genesis of knowledge. The purpose of this section is to briefly review some of the research questions still under-explored in the domain that require community's attention. These questions are not limited to, but just a representative of the amount of gap prevailing in our understanding of the field.

\subsection{Are mediocre people important for knowledge building?}
It has been shown that people make unequal quantitative contributions while building knowledge \cite{de1986little} \cite{muchnik2013origins}. In that context, if only a bunch of people are making most of the contributions, are the people contributing to smaller extents even required in the system? Can we remove them from the system without impeding the growth to a significant extent or are the little contributions made by the mediocre people essential to the system? We still do not have a satisfactory consensus on this fundamental yet critical argument. As an analogy, if coming up with a top invention is like having built a wall, then the question involves finding out whether the wall was completely built by a bunch of top people or several bricks into this wall were added by a large number of people. The forthcoming discussion represents an interesting chronology in the research attempted in this direction. 

The debate over the importance or insignificance dates back to 1932 when Ortega highlighted the importance of common masses in any field \cite{ortega1950revolt}. The same fact has been supported by Florey in \cite{crowther1968science} who asserted that there is nothing called \textit{breakthrough} in science, rather coming up with an excellent piece of work requires small inputs by a large number of people. This hypothesis giving due regard to the small contributions by mediocre people was later termed as \textit{Ortega Hypothesis}. However, Cole and Cole \cite{cole1972ortega} and Bornmann \cite{bornmann2010scientific} are some of the researchers who based on their citation analyses refuted the existence of Ortega Hypothesis. Their studies indicated that top scientists frequently cite the work by other top scientists only and hence the mediocre scientists can be discarded from the system. Therefore they supported the existence of another theory called \textit{Newton Hypothesis} which argues that in any field, most of the work is performed by a small bunch of people and the rest are not needed in the system. However, Hoerman et al. \cite{hoerman1995secondary} pointed out various inaccuracies existing in the citation practices of researchers. Additionally, a detailed study by Macroberts and Macroberts \cite{macroberts1987testing} at the level of checking the cited references found that only less than 30 percent of the references were cited in the studied papers. This clearly indicates that the people or papers that are less famous but do affect the current work, are hardly cited. Moreover, there are a number of not-so-famous people such as those working in labs conducting experiments, which are never cited; hence their importance can not be observed in analyses based on citation analysis. All these works call into question the analyses conducted by Cole and Bornmann. Further, there are two more points worth highlighting here. Firstly, the question of importance of masses is one that is relevant not only in scientific literature, but in any field in general. This is also in accordance with Ortega who highlighted the importance of mediocre people in diverse domains \cite{ortega1950revolt}. Therefore, a general study involving many other domains is required for supporting or refuting the hypothesis. Secondly, if we consider the online knowledge building portals, we observe that mediocre people are playing a crucial role. Wikipedia is the best real-life example demonstrating the power that the masses carry. The success of Wikipedia over experts-based encyclopedia Britannica \cite{giles2005internet} and the failure of all other portals aiming to build knowledge, however, undermining the enormous potential that the crowd possesses, such as \textit{Nupedia}\footnote{https://en.wikipedia.org/wiki/Nupedia}, \textit{Citizendium}\footnote{http://en.citizendium.org} and \textit{Scholarpedia}\footnote{http://www.scholarpedia.org} indicates the fact that the experts alone are unable to successfully build knowledge. Similarly, the enormous success of StackExchange over experts' based Q\&A websites is another indicator of a huge potential of the common people. The success of these portals refutes the existence of Newton Hypothesis. Additionally, Cranshaw and Kittur \cite{cranshaw2011polymath} in their study on Polymath observed that although most of the contribution was made by a small number of contributors, yet, the small amount of content contributed by the rest of the users was very useful for the task under consideration. This further highlights the importance of the little contributions made by the masses. 


The preceding discussion neither supports nor neglects the importance of the common people, rather highlights that the two contradictory schools of thought require a detailed investigation. Getting a clear answer to the question is critical as it may affect the decision-making policies in various fields. As an instance, the decision involving the identification of the right candidates for the distribution of limited government funds for research calls for a clear investigation of the stated dilemma. The answer to this question also forms a basis of the way future knowledge building portals should be built. 

\subsection{How much diversity is important?}
Diversity refers to the differences in peoples' backgrounds, experiences, areas of expertise and cultures etc. Diversity has been proven to be helpful in various situations such as problem solving \cite{hong2004groups} or organizational practices \cite{thomas1996making}. The importance of a diverse crowd has also been emphasized in order to utilize the \textit{wisdom of the crowds} \cite{surowiecki2005wisdom} to the best possible extent. 

It has, however, been shown that a diverse group gives rise to problems such as lack of consensus and more conflicts \cite{chen2010effects}. Further, the success of a diverse crowd is a function of several parameters such as individual ability and group size. For example, there is a connection between diversity and expertise \cite{bonabeau2009decisions}. If the participants have no bare minimum ability, a diverse crowd will not be able to produce good results. Even the mathematical model developed by Hong et al. \cite{hong2004groups} will not hold true if the crowd is not assumed to be having a certain level of ability. Robert et al. \cite{robert2015crowd} studied the relation between crowd size and diversity. They observed that if the crowd size is smaller than a particular threshold, then less diverse teams outperform more diverse teams. It is, therefore, important to study what should be the right amount of diversity in a team given other parameters such as the size of the team and the ability of the participants.

\subsection{Is the knowledge building process convergent?}
As per a classical theory of cognition named \textit{Piaget’s Model of equilibration} \cite{piaget1977development}, people contribute to the knowledge building process because of cognitive conflicts, which means that when they see some information that is incongruent to their existing knowledge, it creates a disturbance in their mind. This disturbance encourages them to add their knowledge to the system. This knowledge addition then leads to the equilibration between the system’s knowledge and the user’s knowledge. However, once the equilibration is reached, no new knowledge comes into the system. Further, the statistics of Wikipedia also indicates a slowing growth of content \cite{suh2009singularity}.  Does it indicate that the knowledge building process always converges? In other words, will we ever reach a stage when we would have possessed all the knowledge that was ever possible to be acquired? However, as already pointed out, a large part of the world's knowledge is yet to be explored \cite{pennisi2003modernizing}. All this discussion points to the need for a thorough analysis into the feasibility of convergence or non-convergence of the process.

\subsection{`More the participants the better' or the law of diminishing returns?}
Surowiecki \cite{surowiecki2005wisdom} in his book, '\textit{Wisdom of crowds}' demonstrated that the aggregate knowledge of a large and diverse group always surpasses that of one or a few experts. This indicates the \textit{more the participants the better} holding true in the knowledge building domain. However, Kittur et al. \cite{kittur2007he} in their study on Wikipedia observed that as the number of participants increases, the amount of conflict among them also increases. This leads to increased co-ordination costs. Therefore, there seems to be a trade-off between the number of participants and the co-ordination amongst them. Further, Robert et al. \cite{robert2015crowd} showed that when the diversity is less, increasing group size does not lead to increase in performance, i.e. in this case, the law of diminishing returns holds. 

On a similar note, in collaborative software development, there are two important laws viz. \textit{Linus Law} \cite{meneely2009secure} and \textit{Brooks Law} \cite{schweik2008brooks}. Linus Law states that `Given enough eyeballs, all bugs are shallow', whereas Brooks law claims that `adding manpower to a late software project makes it even later', i.e. Linus law does not hold true if the project is already delayed. Therefore, more people added to a project actually make it take more time than making it faster. This prompts us to think whether the inclusion of more participants improves the process, or leads to more confusions or it does not matter after a particular limit? Further, is it possible to identify the threshold after which the law of diminishing returns comes into play?

\subsection{Do Participants play different roles?}
A knowledge building system requires its participants to perform diverse activities. For example, in Wikipedia, these could be adding, deleting or updating the text, links, images or references. In a Q\&A portal such as Stack Exchange, these could be questioning, answering or voting etc. Similarly, in the research environment, these activities could be building theories, experimenting, modeling or documenting etc. In that context, do all the participants perform all of these activities or they show inclination towards a few of them? If they do show propensity towards performing a subset of the activities, then what are its implications and does it really affect the system? Further, should a knowledge building system be careful enough to consider this criteria while choosing participants?

Chhabra et al. \cite{chhabra2015presence} observed on a custom crowdsourced annotation system that participants show inclination towards performing mainly one of the activities such as questioning, answering, providing reference or providing insights. Similarly, Ram et al. \cite{liu2011does} identified ten activities performed by contributors on Wikipedia and observed that users perform different roles on the portal constituting a subset of these activities. The same may further be checked on other kinds of systems such as Q\&A portals or collaborative software development etc. Further, It will be interesting to analyze whether the same holds true in the universal knowledge building environment. This analysis may help in encouraging the right mixture of participants for knowledge development.

\subsection{Do people trigger each other?}
As per the information processing theory by Minsky \cite{minsky1977frame}, knowledge is organized into frames. Each of these frames may be considered a unit of knowledge that possesses a particular concept. Hence, these frames may be of varying sizes \cite{fisher1985information}. The frames that are related to each other are linked together \cite{fisher1985information} \cite{norman1981categorization}. Therefore, when a frame is triggered, the frames that are linked to it are also triggered. These frames may be linked sequentially or in any other fashion. One can imagine a forest of nodes where each node is a knowledge frame and the attached frames form the connected components in the forest. Fisher \cite{fisher1985information} further states that these frames are connected by \textit{condition-action} rules, which determine which frames to trigger next. When the triggering conditions for a frame are met, that frame is brought into the system. This leads to a ubiquitous and self-regulating phenomenon of one knowledge unit into the system triggering another unit. Further, Luhmann’s theory \cite{seidl2004luhmann} describes a social system such as a knowledge building system as an \textit{autopoietic system}. This means that once started, further ideas (or cognitions) are produced  by existing ideas of the same system. The existing ideas create perturbations in the cognitive system of users, which then trigger more ideas. The theory further talks about the concept of \textit{structural coupling}, which is the relation between cognitive systems of users and the environment. It states that although the environment can create irritations in all the cognitive systems, it might not be able to trigger all of them. The actual systems that can get triggered due to these perturbations is determined by the \textit{structural coupling} of these systems to the environment. Also, different cognitive systems may have different structural coupling. Although these classical theories indicate that the existing knowledge units trigger more knowledge into the system, they however lack any verification, which poses doubts about their general applicability. It is important to observe the triggering phenomenon - which is one of the important factor behind the creation of new knowledge - into action with the help of underlying data.

\subsection{Is Q\&A necessary for any knowledge building repository?}
The `\textit{Wikipedia}' article on Wikipedia states that it aims to create a summary of all human knowledge \footnote{https://en.wikipedia.org/wiki/Wikipedia}. However, from past few years, it has been showing signs of slowing growth  \cite{suh2009singularity}. It will not be right to assume that Wikipedia has already covered most of the knowledge that humans possess today. It is important for the readers to convey the Wikipedia article editors of the gaps prevailing in the articles. Therefore, it is important to have a communication medium between the readers and the editors of the repository. For that, we believe that inclusion of a Q\&A facility on Wikipedia might let the volunteer editors know of the gaps that exist in the encyclopedia currently. In general, by taking an example of Wikipedia, we hypothesize that Q\&A should be an integral part of every knowledge building repository for better knowledge creation \cite{chhabra2016should}.  

Moreover, not only knowledge repositories might benefit by incorporating Q\&A, rather, the same holds true the other way around as well. In any Q\&A portal, the knowledge is scattered in a non-uniform fashion across various threads. Preparing a repository out of the knowledge gathered from the questions and answers will be helpful in keeping related concepts at one place. In that regard, StackExchange recently has started the \textit{documentation initiative}\footnote{http://stackoverflow.com/documentation} to build a repository of knowledge out of questions and answers.

\subsection{Does incentivization help?}
Incentivization is known to improve user participation and efficiency in any system \cite{anderson2013steering}. Several Q\&A portals such as StackExchange and Quora use various incentivization techniques by way of votes, points and badges etc and are successfully utilizing crowds' potential. However, incentivization in another successful portal, i.e. Wikipedia, is limited to mere providing some administrative privileges to the users contributing actively. Despite that, Wikipedia has been quite successful in eliciting knowledge from the masses. If it introduces points and badges like other portals, will it perform better or it will discourage participants following \textit{Pay enough or don't pay at all} policy \cite{gneezy2000pay}? The effects of incentivization in the scientific knowledge building is another interesting work that may provide further insights, i.e., is there anyway we can observe the effects of incentivization in the human knowledge building process?

The preceding discussion suggests that knowledge building is affected by many parameters, where some of them may catalyze the process while some others may be detrimental to the process. It is important to have a clear understanding of these parameters to be able to discern the phenomenon of building new knowledge. In the next section, we propose to  follow a four-step procedure to comprehend the knowledge building aspects in a better way.

\section{Our Vision: The AADE Architecture}
\begin{figure*}
\centering
\includegraphics[scale = 0.29]{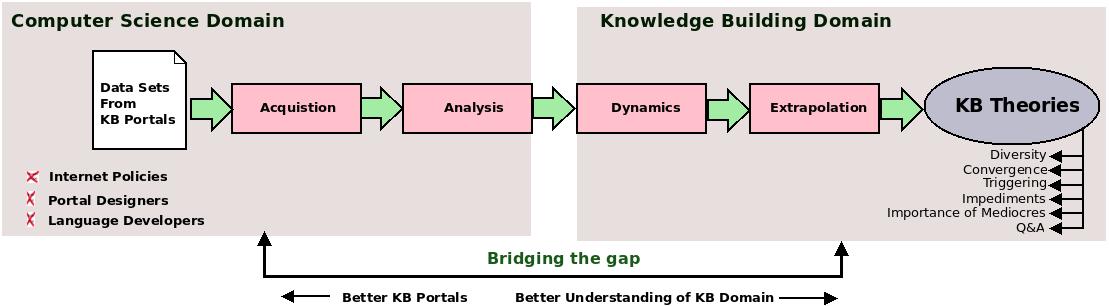}
\caption{AADE Architecture with the Acquisition, Analysis, Dynamics and Extrapolation Phases (*KB : Knowledge Building)}
\label{AADE}
\end{figure*}

We envision an in-depth analysis of the data sets availed from online portals, which are currently under-explored in terms of understanding the intricate details of the knowledge building process. The insights obtained from the analyses show a great promise of both settling some of the long-pending debates in the area as well as a better know-how of the process. For that, we need to first acquire the underlying data, perform various analyses on the data, identify the various factors affecting the process and then extrapolate these insights to decipher the human knowledge building phenomenon. Based on these activities, we propose an architecture termed as \textit{AADE Architecture} (See Figure~\ref{AADE}) consisting of four phases, as listed below:
\begin{enumerate}
\item Acquisition (A)
\item Analysis (A)
\item Dynamics (D)
\item Extrapolate (E)
\end{enumerate}

\subsection{Acquisition} The first step is to acquire the data sets from the knowledge building portals. Some portals provide the underlying data openly available for public in various formats such as XML or SQL. For example, StackExchange and Wikipedia provide all their data sets periodically for the research community to access and use, whereas, others like Quora and Polymath do not release their underlying data. Therefore, this phase may require use of techniques such as crawling or development of APIs for accessing the data sets. Accessing data may also require authorization and facilitation of the appropriate infrastructure. For example, Wikipedia data set is as big as several tera bytes. In order to utilize this extensive data, enough memory and processing capabilities might be needed. 

Due to developments in Web 2.0, there are several crowdsourced portals enabling large scale collaboration. Table~\ref{portals} shows some of the candidate portals offering diverse features and capturing different dimensions of the process. Below we provide a brief introduction to these portals.

\begin{table}
\centering
\begin{tabular}{|p{2cm}|p{1.5cm}|p{2.5cm}|}
\hline
\textbf{Name} & \textbf{Founding year} & \textbf{Type}\\
\hline
\rule{0pt}{2ex}   Wikipedia & 2001 & Encyclopedia\\
\hline
\rule{0pt}{2ex}   StackExchange & 2008 & Q\&A \\
\hline
\rule{0pt}{2ex}   Quora & 2009 & Q\&A\\
\hline
\rule{0pt}{2ex}   YahooAnswers &2005 & Q\&A\\
\hline
\rule{0pt}{2ex}   Polymath &2009 & Collaborative Theorem Proving\\
\hline
\rule{0pt}{2ex}   Github & 2008& Collaborative Software Development\\
\hline
\end{tabular}
\caption{Some Crowdsourced Knowledge Building Portals}
\label{portals}
\end{table}

\textit{\textbf{Wikipedia}} is the best attempt in eliciting knowledge possessed by individuals from diverse, demographic segments of the society. These individuals voluntarily collaborate on various topics and provide their share of knowledge. Currently enjoying an alexa rank of 6, this portal has 5,341,019 content pages with 30,283,660 registered users with the numbers increasing every day. It is the best example of collaboration among such a huge number of people over a diverse range of topics. Wikimedia foundation makes Wikipedia's data dumps freely available for use by public, twice every month \footnote{https://dumps.wikimedia.org/enwiki/}. These dumps are available in both XML as well as SQL. 

\textit{\textbf{StackExchange}} which started with a single programming based Q\&A website, StackOverflow, currently has a collection of around 165 websites on technical as well non-technical fact-based genres. Although the initial aim of the portal was to help people find solutions to their queries, it has in turn built an extensive and highly useful repository of knowledge in the form of questions and answers. The statistics indicate 7.9 billion pageviews, 3.7 million question asked and 4.6 million answers submitted. StackExchange too provides its data sets periodically in the form of XML files for analysis.

\textit{\textbf{Quora}} is another successful portal that augments a Q\&A system with social links between users. This portal helps people seek and provide perspective based answers, hence building another very useful repository of knowledge. People come here not only to get technical assistance but also to solicit advice, opinions and satisfy their curiosity about various things in life. Due to no publicly available data sets of Quora, we come across very limited work on this portal that uses crawlers to obtain the underlying details.

\textit{\textbf{Yahoo! Answers}} is one of the initial Q\&A websites. Although it was once topping the charts due to its success \cite{mcgee2010yahoo}, it has currently almost stalled and has declined in terms of growth. Nevertheless, it is a good candidate for analysis of the factors that may be detrimental to the growth of knowledge. That is, a live portal that was once on peak and was gradually sidelined by the introduction of other portals is all the more important to study for research reasons. It may help identifying the parameters that may hinder the proliferation of knowledge in an environment.

\textit{\textbf{Polymath Project}} is a portal for collaboration among mathematicians to collectively solve important and difficult mathematical problems. It is one of the successful examples of collaboration among professional scientists for true discoveries. Although the scientists have been collaborating in the past while they came up with new discoveries, it was never possible to access the traces of these collaborations and hence to study them. This portal is however a perfect example to witness the collaboration taking place online. Therefore, an access to its data may help in improving the collaboration dynamics of the scientific knowledge building in general. 


\textit{\textbf{Github}} is a version control system offering various functionalities such as bug tracking, feature requests, task management and wikis for building software collaboratively. Software development is an area where collaboration is in fact difficult to achieve and there is high probability of conflicts. It requires use of efficient coordination among the contributors. Github has proven itself as a medium for collaboration over extensive software projects with its extensive success. Therefore, the data set taken from this portal can help in formalizing theories regarding collaboration in general.

\subsection{Analysis} The next step is to perform analyses on the data sets. Different portals focus on different aspects of the process. Hence, it is important to choose the right portal and hence the data set depending on which aspect of the knowledge building process we wish to analyze. For example, for analyzing the effect of incentivization, StackExchange will be the right portal than Wikipedia, whereas the later may be a superior playground to observe large scale collaboration over an artifact of knowledge. Further, this phase may require performing large scale as well as in-depth analyses on the data sets. Therefore, programming languages such as Python, Matlab etc, various APIs, and big data tools such as Hadoop infrastructure will be used. 

\subsection{Dynamics}
Based on the analyses of the data sets, the next step is to understand the dynamics of the knowledge building process. In particular, the acquisition and analysis are the phases where computer science researchers will be of great help, whereas the dynamics phase requires acquaintance with the existing knowledge building confusions and a know-how of the related cognitive theories. This phase will help in finding the answers to the questions stated in the section 3 on the knowledge building conundrum. We see a frequent to-and-fro flow of information between the analysis and dynamics phase. For example, verifying the existence of triggering might require making sense out of the word-level analyses conducted on the edit histories of Wikipedia articles, indicating which factoid might have triggered the inclusion of which other factoid. Further, based on what we understand, the analyses might be modified.

\subsection{Extrapolate}
This phase requires the verification of the theories, that we built out of the analyses of data sets, in the universal knowledge building environment. It may happen that some of the theories that we may build in the dynamics phase apply to a subset of people or situations. Therefore, this phase extrapolates the insights obtained and checks for their general applicability.

\section{Challenges and Future Directions}
The current scenario might not be suitable to efficiently execute the proposed phases. We foresee a number of difficulties while working in various phases of the architecture. In this section we highlight some of the key challenges that must be addressed to employ use of online portals' data sets to establish knowledge building concepts.

\subsection{Data Acquisition Challenges}
In order to realize the aim, it is imperative to find out or devise easy methods to acquire data sets. A huge amount of overhead in accessing the data may deter the knowledge building researchers from using them. On contrary, provision of data sets in the right format will enable their quick and easy use by the researchers with limited resources as well. We underline some of the currently existing challenges here. For various reasons, some of the portals such as Quora, Google Scholar do not release their underlying data sets. The only option to access the underlying data of the portals which do not release their data is to use crawlers, however, there are various restrictions imposed by the portals in terms of how much and how long can they access data. Hence, using such methods leads to a very limited analysis. To resolve this, Internet standards should be such that it should be mandatory for every crowdsourced portal seeking the services of the crowd to keep its data open for analysis at regular intervals. The data may optionally be \textit{anonymized} \cite{sweeney1997weaving} if needed. Moreover, these data sets should conform to a common set of guidelines for a better analysis. These guidelines could include using the standard data storage formats such as JSON, XML etc. Further, the portals should provide their data in easy to handle chunks, where each chunk carries a different kind of information. This would reduce the requirement of extensive memory and other resources. For instance, although Wikipedia provides its data dumps periodically, these dumps are enormous in size, which is no surprise, given the size of the encyclopedia. However, there are certain issues while using them in an infrastructure with limited memory and processing capabilities. For example, in the detailed data set, the revisions', talk pages', users' etc details about one particular article are scattered over a number of xml files, which makes it necessary for a user to download the entire dump and perform their analysis over it. There are various ways that the Wikimedia foundation may employ to provide their data dumps on a per article basis or any other easy-to-trace format so that downloading a subset of the dump is enough for the analyses that do not require the whole dump. 

Further, acquiring the right kind of data may also require development of new infrastructure if none of the existing portals is able to provide the required data set. For example, checking whether Q\&A is a necessary feature of any knowledge repository and vice versa or not \cite{chhabra2016should}, would require creation of new portals exploiting this feature or incorporation of new features in the existing portals. This is due to the fact that none of the existing portals currently validates or refutes this aspect. Similarly, Chhabra et al. \cite{chhabra2015presence} created a custom portal based on annotations named\textit{CAS} to observe whether participants exhibit inclination towards performing mostly one of the activities on an crowdsourced annotation portal.

\subsection{Data Analysis Challenges}
Firstly, there are no APIs for many of the portals. Secondly, there are a few API's for analysis of data from Wikipedia, such as \textit{Wikipedia, Pywikipediabot} and \textit{wiki-api}, and for StackExchange, we have StackAPI and Py-StackExchange, however, they are very limited in their functionalities. Other than providing summaries or some basic statistics, they can not be used for any advanced analyses. For example, if we wish to thoroughly analyze the revision histories and compare the successive edits made to a given article of Wikipedia, there is no quick methods or tools. To overcome this limitation, language developers can contribute by identifying the commonly occurring functions that are usually needed for analyzing the portals and hence building extensive APIs for these functions. This will further promote the use of these data sets by non computer science researchers without requiring them to get into the low level details.

\subsection{Challenge of Multidisciplinarity}
As highlighted earlier, out of the four phases of the AADE architecture, the first two require working with computer science tools and technologies for accessing and analyzing the data sets, while the later two phases involve working with the existing theories of the knowledge building domain. The gaps in the understanding of the knowledge building domain are known by the knowledge building community. However, they have very limited knowledge of handling big data. On the other hand, computer science community is well versed with the tools and techniques for handling the huge amount of data, but are unaware of the kinds of analyses that are required for the advancements of knowledge community. The current scenario indicates very limited collaboration between scientists from these disciplines. For instance, computer scientists are focusing on conducting studies to understand the knowledge building portals. Knowledge building researchers are focusing on building theories based on very small scale observations without their general applicability. It should be noted that a collaboration between the researchers from these two domains might benefit both the domains in terms of providing a better idea of the knowledge building process as well as the necessary insights for designing better portals. Further, deciphering various aspects of the process is an interdisciplinary work requiring assistance from cognitive and social science domains as well. Therefore, a collective effort by researchers from all these domains will be instrumental in realizing our aim.

\section{Discussion}
It would not be an overstatement if we proclaim that the area of knowledge building is one of the most important areas for the research community. This is because it can help in explaining what leads to the discovery of new things or what poses a hindrance to it. Therefore, improvements in this domain might in turn foster consequential advancements in other domains. However several confusions pertaining to the domain indicate that we do not understand it well enough. For example, we ask a very fundamental and simple question: \textit{How does new knowledge come by?} Does it happen by chance? Or there are some factors that lead to the discovery of new facts? As an instance, if, somehow, the civilization is demolished with only a small bunch of people alive, with no past record of knowledge, how many years will it take for those bunch of people to reach the stage that we are currently in? The research in the field till date is not able to answer this very basic question. We simply do not know what leads to the synthesis of new knowledge. This, along with a number of other uncertainties in the domain indicates a huge lack of our understanding of the field.

To be able to fill these gaps, the first step is to understand the process along with its inherent characteristics. Without a proper understanding of the phenomenon, the scope of improvements in this domain stays dependent only on chance happenings. The primary reason that we have not been able to carry out significant research in the domain is the lack of data that may provide us the underlying details. However, with the advancements in the Internet technologies, current times pose a great deal of promise in this direction by facilitating portals enabling us to build knowledge collaboratively. An added advantage is that it is possible to store the details of the activities taking place over these portals in the form of data sets. These data sets which may provide us the footprints of knowledge built over the portal, may help us observe what lead to its creation. Extrapolating these insights for the universal knowledge building, we may be able to answer the general question of how new knowledge gets built. The huge volumes of data that are available over these portals are a contribution of the recent times only and was not available a few years back. In the absence of this data, the community had to rely on the theoretical or small-scale experiments only to `estimate' what was happening in the domain. Therefore, it is high time knowledge building community starts making use of this wealth of meta-knowledge to gain insights into the process. The wisdom thus gained may help in both developing new theories as well as validating the existing ones. This in turn will help in further advancing the domain and hence in swift unearthing a large number of things that we are yet to explore.

There are, however, several challenges such as non-disclosure of data by some portals, difficulty in accessing the right kind of data and non-availability of tools to ease up the understanding of a huge corpus of data sets. For example, currently, we find very limited APIs and tools for the analysis of huge data sets from portals like Wikipedia, StackExchange etc and no tools for other crowdsourced portals. This, in turn, hinders their use by knowledge building researchers with limited knowledge of handling big data. To enable easy access to these footprints, several measures can be taken by the Internet community, portals designers and the language developers. For example, disclosing the underlying data sets in some standard formats should be made mandatory for any portal making use of the services of the crowd. 

\section{Conclusion}
The aim of this report is not to encourage the \textit{computer science community} to make use of the data sets from crowdsourced portals, which they are already doing, mostly for portal specific analyses. We wish to rather bring the attention of the \textit{knowledge building community} towards this huge meta-knowledge in order to answer a broad question as to what factors lead to the creation of new knowledge. Apart from answering this question, the analyses conducted on this data will have many-fold benefits. A better understanding of the associated concepts will help in confirming the disputed theoretical aspects of the domain. Further, knowing about the catalysts and the impediments of the process, the portal designers will be able to build conducive environments to foster better knowledge building. Moreover, the administrators, if well equipped with the knowledge of an ideal knowledge building environment, may take informed measures whenever they observe a lack of growth. 
\bibliographystyle{IEEEtran}
\bibliography{KB_bibliography}

\end{document}